\documentclass[10pt,conference,twoside]{IEEEtran}
\usepackage{cite}
\usepackage[T1]{fontenc}
\usepackage{graphicx}
\usepackage{amssymb}
\usepackage{amsmath}
\usepackage{paralist}
\usepackage{subfigure}
\usepackage{booktabs} 
\usepackage{algorithm}
\usepackage{algorithmic}
\usepackage{amsthm}
\usepackage{multirow}
\usepackage{microtype}
\usepackage{tablefootnote}

\usepackage{amssymb}
\usepackage{amsfonts}
\usepackage{mathrsfs}
\usepackage{xspace}
\usepackage{bm}
\usepackage{upgreek}

\newcommand{\safemath}[2]{\newcommand{#1}{\ensuremath{#2}\xspace}}

\safemath{\bma}{\mathbf{a}}
\safemath{\bmb}{\mathbf{b}}
\safemath{\bmc}{\mathbf{c}}
\safemath{\bmd}{\mathbf{d}}
\safemath{\bme}{\mathbf{e}}
\safemath{\bmf}{\mathbf{f}}
\safemath{\bmg}{\mathbf{g}}
\safemath{\bmh}{\mathbf{h}}
\safemath{\bmi}{\mathbf{i}}
\safemath{\bmj}{\mathbf{j}}
\safemath{\bmk}{\mathbf{k}}
\safemath{\bml}{\mathbf{l}}
\safemath{\bmm}{\mathbf{m}}
\safemath{\bmn}{\mathbf{n}}
\safemath{\bmo}{\mathbf{o}}
\safemath{\bmp}{\mathbf{p}}
\safemath{\bmq}{\mathbf{q}}
\safemath{\bmr}{\mathbf{r}}
\safemath{\bms}{\mathbf{s}}
\safemath{\bmt}{\mathbf{t}}
\safemath{\bmu}{\mathbf{u}}
\safemath{\bmv}{\mathbf{v}}
\safemath{\bmw}{\mathbf{w}}
\safemath{\bmx}{\mathbf{x}}
\safemath{\bmy}{\mathbf{y}}
\safemath{\bmz}{\mathbf{z}}
\safemath{\bmzero}{\mathbf{0}}
\safemath{\bmone}{\mathbf{1}}

\bmdefine{\biad}{a}
\bmdefine{\bibd}{b}
\bmdefine{\bicd}{c}
\bmdefine{\bidd}{d}
\bmdefine{\bied}{e}
\bmdefine{\bifd}{f}
\bmdefine{\bigd}{g}
\bmdefine{\bihd}{h}
\bmdefine{\biid}{i}
\bmdefine{\bijd}{j}
\bmdefine{\bikd}{k}
\bmdefine{\bild}{l}
\bmdefine{\bimd}{m}
\bmdefine{\bind}{n}
\bmdefine{\biod}{o}
\bmdefine{\bipd}{p}
\bmdefine{\biqd}{q}
\bmdefine{\bird}{r}
\bmdefine{\bisd}{s}
\bmdefine{\bitd}{t}
\bmdefine{\biud}{u}
\bmdefine{\bivd}{v}
\bmdefine{\biwd}{w}
\bmdefine{\bixd}{x}
\bmdefine{\biyd}{y}
\bmdefine{\bizd}{z}

\bmdefine{\bixid}{\xi}
\bmdefine{\bilambdad}{\lambda}
\bmdefine{\bimud}{\mu}
\bmdefine{\bithetad}{\theta}
\bmdefine{\biphid}{\phi}
\bmdefine{\bideltad}{\delta}

\safemath{\bmia}{\biad}
\safemath{\bmib}{\bibd}
\safemath{\bmic}{\bicd}
\safemath{\bmid}{\bidd}
\safemath{\bmie}{\bied}
\safemath{\bmif}{\bifd}
\safemath{\bmig}{\bigd}
\safemath{\bmih}{\bihd}
\safemath{\bmii}{\biid}
\safemath{\bmij}{\bijd}
\safemath{\bmik}{\bikd}
\safemath{\bmil}{\bild}
\safemath{\bmim}{\bimd}
\safemath{\bmin}{\bind}
\safemath{\bmio}{\biod}
\safemath{\bmip}{\bipd}
\safemath{\bmiq}{\biqd}
\safemath{\bmir}{\bird}
\safemath{\bmis}{\bisd}
\safemath{\bmit}{\bitd}
\safemath{\bmiu}{\biud}
\safemath{\bmiv}{\bivd}
\safemath{\bmiw}{\biwd}
\safemath{\bmix}{\bixd}
\safemath{\bmiy}{\biyd}
\safemath{\bmiz}{\bizd}

\safemath{\bmxi}{\bixid}
\safemath{\bmlambda}{\bilambdad}
\safemath{\bmmu}{\bimud}
\safemath{\bmtheta}{\bithetad}
\safemath{\bmphi}{\biphid}
\safemath{\bmdelta}{\bideltad}

\safemath{\bA}{\mathbf{A}}
\safemath{\bB}{\mathbf{B}}
\safemath{\bC}{\mathbf{C}}
\safemath{\bD}{\mathbf{D}}
\safemath{\bE}{\mathbf{E}}
\safemath{\bF}{\mathbf{F}}
\safemath{\bG}{\mathbf{G}}
\safemath{\bH}{\mathbf{H}}
\safemath{\bI}{\mathbf{I}}
\safemath{\bJ}{\mathbf{J}}
\safemath{\bK}{\mathbf{K}}
\safemath{\bL}{\mathbf{L}}
\safemath{\bM}{\mathbf{M}}
\safemath{\bN}{\mathbf{N}}
\safemath{\bO}{\mathbf{O}}
\safemath{\bP}{\mathbf{P}}
\safemath{\bQ}{\mathbf{Q}}
\safemath{\bR}{\mathbf{R}}
\safemath{\bS}{\mathbf{S}}
\safemath{\bT}{\mathbf{T}}
\safemath{\bU}{\mathbf{U}}
\safemath{\bV}{\mathbf{V}}
\safemath{\bW}{\mathbf{W}}
\safemath{\bX}{\mathbf{X}}
\safemath{\bY}{\mathbf{Y}}
\safemath{\bZ}{\mathbf{Z}}

\safemath{\bZero}{\mathbf{0}}
\safemath{\bOne}{\mathbf{1}}
\safemath{\bDelta}{\mathbf{\Delta}}
\safemath{\bLambda}{\mathbf{\UpLambda}}
\safemath{\bPhi}{\mathbf{\Upphi}}
\safemath{\bSigma}{\mathbf{\Upsigma}}
\safemath{\bOmega}{\mathbf{\Upomega}}
\safemath{\bTheta}{\mathbf{\Uptheta}}

\bmdefine{\biAd}{A}
\bmdefine{\biBd}{B}
\bmdefine{\biCd}{C}
\bmdefine{\biDd}{D}
\bmdefine{\biEd}{E}
\bmdefine{\biFd}{F}
\bmdefine{\biGd}{G}
\bmdefine{\biHd}{H}
\bmdefine{\biId}{I}
\bmdefine{\biJd}{J}
\bmdefine{\biKd}{K}
\bmdefine{\biLd}{L}
\bmdefine{\biMd}{M}
\bmdefine{\biOd}{N}
\bmdefine{\biPd}{O}
\bmdefine{\biQd}{P}
\bmdefine{\biRd}{R}
\bmdefine{\biSd}{S}
\bmdefine{\biTd}{T}
\bmdefine{\biUd}{U}
\bmdefine{\biVd}{V}
\bmdefine{\biWd}{W}
\bmdefine{\biXd}{X}
\bmdefine{\biYd}{Y}
\bmdefine{\biZd}{Z}

\bmdefine{\biDelta}{\Delta}
\bmdefine{\biLambda}{\Lambda}
\bmdefine{\biPhi}{\Phi}
\bmdefine{\biSigma}{\Sigma}
\bmdefine{\biOmega}{\Omega}
\bmdefine{\biTheta}{\Theta}

\safemath{\bimA}{\biAd}
\safemath{\bimB}{\biBd}
\safemath{\bimC}{\biCd}
\safemath{\bimD}{\biDd}
\safemath{\bimE}{\biEd}
\safemath{\bimF}{\biFd}
\safemath{\bimG}{\biGd}
\safemath{\bimH}{\biHd}
\safemath{\bimI}{\biId}
\safemath{\bimJ}{\biJd}
\safemath{\bimK}{\biKd}
\safemath{\bimL}{\biLd}
\safemath{\bimM}{\biMd}
\safemath{\bimN}{\biNd}
\safemath{\bimO}{\biOd}
\safemath{\bimP}{\biPd}
\safemath{\bimQ}{\biQd}
\safemath{\bimR}{\biRd}
\safemath{\bimS}{\biSd}
\safemath{\bimT}{\biTd}
\safemath{\bimU}{\biUd}
\safemath{\bimV}{\biVd}
\safemath{\bimW}{\biWd}
\safemath{\bimX}{\biXd}
\safemath{\bimY}{\biYd}
\safemath{\bimZ}{\biZd}

\safemath{\bimDelta}{\biDelta}
\safemath{\bimLambda}{\biLambda}
\safemath{\bimPhi}{\biPhi}
\safemath{\bimSigma}{\biSigma}
\safemath{\bimOmega}{\biOmega}
\safemath{\bimTheta}{\biTheta}

\safemath{\setA}{\mathcal{A}}
\safemath{\setB}{\mathcal{B}}
\safemath{\setC}{\mathcal{C}}
\safemath{\setD}{\mathcal{D}}
\safemath{\setE}{\mathcal{E}}
\safemath{\setF}{\mathcal{F}}
\safemath{\setG}{\mathcal{G}}
\safemath{\setH}{\mathcal{H}}
\safemath{\setI}{\mathcal{I}}
\safemath{\setJ}{\mathcal{J}}
\safemath{\setK}{\mathcal{K}}
\safemath{\setL}{\mathcal{L}}
\safemath{\setM}{\mathcal{M}}
\safemath{\setN}{\mathcal{N}}
\safemath{\setO}{\mathcal{O}}
\safemath{\setP}{\mathcal{P}}
\safemath{\setQ}{\mathcal{Q}}
\safemath{\setR}{\mathcal{R}}
\safemath{\setS}{\mathcal{S}}
\safemath{\setT}{\mathcal{T}}
\safemath{\setU}{\mathcal{U}}
\safemath{\setV}{\mathcal{V}}
\safemath{\setW}{\mathcal{W}}
\safemath{\setX}{\mathcal{X}}
\safemath{\setY}{\mathcal{Y}}
\safemath{\setZ}{\mathcal{Z}}
\safemath{\emptySet}{\varnothing}

\safemath{\colA}{\mathscr{A}}
\safemath{\colB}{\mathscr{B}}
\safemath{\colC}{\mathscr{C}}
\safemath{\colD}{\mathscr{D}}
\safemath{\colE}{\mathscr{E}}
\safemath{\colF}{\mathscr{F}}
\safemath{\colG}{\mathscr{G}}
\safemath{\colH}{\mathscr{H}}
\safemath{\colI}{\mathscr{I}}
\safemath{\colJ}{\mathscr{J}}
\safemath{\colK}{\mathscr{K}}
\safemath{\colL}{\mathscr{L}}
\safemath{\colM}{\mathscr{M}}
\safemath{\colN}{\mathscr{N}}
\safemath{\colO}{\mathscr{O}}
\safemath{\colP}{\mathscr{P}}
\safemath{\colQ}{\mathscr{Q}}
\safemath{\colR}{\mathscr{R}}
\safemath{\colS}{\mathscr{S}}
\safemath{\colT}{\mathscr{T}}
\safemath{\colU}{\mathscr{U}}
\safemath{\colV}{\mathscr{V}}
\safemath{\colW}{\mathscr{W}}
\safemath{\colX}{\mathscr{X}}
\safemath{\colY}{\mathscr{Y}}
\safemath{\colZ}{\mathscr{Z}}

\safemath{\opA}{\mathbb{A}}
\safemath{\opB}{\mathbb{B}}
\safemath{\opC}{\mathbb{C}}
\safemath{\opD}{\mathbb{D}}
\safemath{\opE}{\mathbb{E}}
\safemath{\opF}{\mathbb{F}}
\safemath{\opG}{\mathbb{G}}
\safemath{\opH}{\mathbb{H}}
\safemath{\opI}{\mathbb{I}}
\safemath{\opJ}{\mathbb{J}}
\safemath{\opK}{\mathbb{K}}
\safemath{\opL}{\mathbb{L}}
\safemath{\opM}{\mathbb{M}}
\safemath{\opN}{\mathbb{N}}
\safemath{\opO}{\mathbb{O}}
\safemath{\opP}{\mathbb{P}}
\safemath{\opQ}{\mathbb{Q}}
\safemath{\opR}{\mathbb{R}}
\safemath{\opS}{\mathbb{S}}
\safemath{\opT}{\mathbb{T}}
\safemath{\opU}{\mathbb{U}}
\safemath{\opV}{\mathbb{V}}
\safemath{\opW}{\mathbb{W}}
\safemath{\opX}{\mathbb{X}}
\safemath{\opY}{\mathbb{Y}}
\safemath{\opZ}{\mathbb{Z}}
\safemath{\opZero}{\mathbb{O}}
\safemath{\identityop}{\opI}

\safemath{\veca}{\bma}
\safemath{\vecb}{\bmb}
\safemath{\vecc}{\bmc}
\safemath{\vecd}{\bmd}
\safemath{\vece}{\bme}
\safemath{\vecf}{\bmf}
\safemath{\vecg}{\bmg}
\safemath{\vech}{\bmh}
\safemath{\veci}{\bmi}
\safemath{\vecj}{\bmj}
\safemath{\veck}{\bmk}
\safemath{\vecl}{\bml}
\safemath{\vecm}{\bmm}
\safemath{\vecn}{\bmn}
\safemath{\veco}{\bmo}
\safemath{\vecp}{\bmp}
\safemath{\vecq}{\bmq}
\safemath{\vecr}{\bmr}
\safemath{\vecs}{\bms}
\safemath{\vect}{\bmt}
\safemath{\vecu}{\bmu}
\safemath{\vecv}{\bmv}
\safemath{\vecw}{\bmw}
\safemath{\vecx}{\bmx}
\safemath{\vecy}{\bmy}
\safemath{\vecz}{\bmz}

\safemath{\veczero}{\bmzero}
\safemath{\vecone}{\bmone}
\safemath{\vecxi}{\bmxi}
\safemath{\veclambda}{\bmlambda}
\safemath{\vecmu}{\bmmu}
\safemath{\vectheta}{\bmtheta}
\safemath{\vecphi}{\bmphi}
\safemath{\vecdelta}{\bmdelta}

\safemath{\matA}{\bA}
\safemath{\matB}{\bB}
\safemath{\matC}{\bC}
\safemath{\matD}{\bD}
\safemath{\matE}{\bE}
\safemath{\matF}{\bF}
\safemath{\matG}{\bG}
\safemath{\matH}{\bH}
\safemath{\matI}{\bI}
\safemath{\matJ}{\bJ}
\safemath{\matK}{\bK}
\safemath{\matL}{\bL}
\safemath{\matM}{\bM}
\safemath{\matN}{\bN}
\safemath{\matO}{\bO}
\safemath{\matP}{\bP}
\safemath{\matQ}{\bQ}
\safemath{\matR}{\bR}
\safemath{\matS}{\bS}
\safemath{\matT}{\bT}
\safemath{\matU}{\bU}
\safemath{\matV}{\bV}
\safemath{\matW}{\bW}
\safemath{\matX}{\bX}
\safemath{\matY}{\bY}
\safemath{\matZ}{\bZ}
\safemath{\matzero}{\bmzero}

\safemath{\matDelta}{\bDelta}
\safemath{\matLambda}{\bLambda}
\safemath{\matPhi}{\bPhi}
\safemath{\matSigma}{\bSigma}
\safemath{\matOmega}{\bOmega}
\safemath{\matTheta}{\bTheta}

\safemath{\matidentity}{\matI}
\safemath{\matone}{\matO}

\safemath{\rnda}{A}
\safemath{\rndb}{B}
\safemath{\rndc}{C}
\safemath{\rndd}{D}
\safemath{\rnde}{E}
\safemath{\rndf}{F}
\safemath{\rndg}{G}
\safemath{\rndh}{H}
\safemath{\rndi}{I}
\safemath{\rndj}{J}
\safemath{\rndk}{K}
\safemath{\rndl}{L}
\safemath{\rndm}{M}
\safemath{\rndn}{N}
\safemath{\rndo}{O}
\safemath{\rndp}{P}
\safemath{\rndq}{Q}
\safemath{\rndr}{R}
\safemath{\rnds}{S}
\safemath{\rndt}{T}
\safemath{\rndu}{U}
\safemath{\rndv}{V}
\safemath{\rndw}{W}
\safemath{\rndx}{X}
\safemath{\rndy}{Y}
\safemath{\rndz}{Z}

\safemath{\rveca}{\bimA}
\safemath{\rvecb}{\bimB}
\safemath{\rvecc}{\bimC}
\safemath{\rvecd}{\bimD}
\safemath{\rvece}{\bimE}
\safemath{\rvecf}{\bimF}
\safemath{\rvecg}{\bimG}
\safemath{\rvech}{\bimH}
\safemath{\rveci}{\bimI}
\safemath{\rvecj}{\bimJ}
\safemath{\rveck}{\bimK}
\safemath{\rvecl}{\bimL}
\safemath{\rvecm}{\bimM}
\safemath{\rvecn}{\bimN}
\safemath{\rveco}{\bomO}
\safemath{\rvecp}{\bimP}
\safemath{\rvecq}{\bimQ}
\safemath{\rvecr}{\bimR}
\safemath{\rvecs}{\bimS}
\safemath{\rvect}{\bimT}
\safemath{\rvecu}{\bimU}
\safemath{\rvecv}{\bimV}
\safemath{\rvecw}{\bimW}
\safemath{\rvecx}{\bimX}
\safemath{\rvecy}{\bimY}
\safemath{\rvecz}{\bimZ}

\safemath{\rvecxi}{\bmxi}
\safemath{\rveclambda}{\bmlambda}
\safemath{\rvecmu}{\bmmu}
\safemath{\rvectheta}{\bmtheta}
\safemath{\rvecphi}{\bmphi}

\safemath{\rmatA}{\bimA}
\safemath{\rmatB}{\bimB}
\safemath{\rmatC}{\bimC}
\safemath{\rmatD}{\bimD}
\safemath{\rmatE}{\bimE}
\safemath{\rmatF}{\bimF}
\safemath{\rmatG}{\bimG}
\safemath{\rmatH}{\bimH}
\safemath{\rmatI}{\bimI}
\safemath{\rmatJ}{\bimJ}
\safemath{\rmatK}{\bimK}
\safemath{\rmatL}{\bimL}
\safemath{\rmatM}{\bimM}
\safemath{\rmatN}{\bimN}
\safemath{\rmatO}{\bimO}
\safemath{\rmatP}{\bimP}
\safemath{\rmatQ}{\bimQ}
\safemath{\rmatR}{\bimR}
\safemath{\rmatS}{\bimS}
\safemath{\rmatT}{\bimT}
\safemath{\rmatU}{\bimU}
\safemath{\rmatV}{\bimV}
\safemath{\rmatW}{\bimW}
\safemath{\rmatX}{\bimX}
\safemath{\rmatY}{\bimY}
\safemath{\rmatZ}{\bimZ}

\safemath{\rmatDelta}{\bimDelta}
\safemath{\rmatLambda}{\bimLambda}
\safemath{\rmatPhi}{\bimPhi}
\safemath{\rmatSigma}{\bimSigma}
\safemath{\rmatOmega}{\bimOmega}
\safemath{\rmatTheta}{\bimTheta}

\usepackage{amssymb}
\usepackage{amsfonts}
\usepackage{mathrsfs}
\usepackage{xspace}
\usepackage{bm}
\usepackage{fancyref}
\usepackage{textcomp}

\usepackage{multirow}
\usepackage{stmaryrd}

\newenvironment{textbmatrix}{	\setlength{\arraycolsep}{2.5pt}%
								\big[\begin{matrix}}{\end{matrix}\big]%
								\raisebox{0.08ex}{\vphantom{M}}}

\def\be{\begin{equation}}
\def\ee{\end{equation}}
\def\een{\nonumber \end{equation}}
\def\mat{\begin{bmatrix}}
\def\emat{\end{bmatrix}}
\def\btm{\begin{textbmatrix}}
\def\etm{\end{textbmatrix}}

\def\ba#1\ea{\begin{align}#1\end{align}}
\def\bas#1\eas{\begin{align*}#1\end{align*}}
\def\bs#1\es{\begin{split}#1\end{split}}
\def\bg#1\eg{\begin{gather}#1\end{gather}}
\def\bml#1\eml{\begin{multline}#1\end{multline}}
\def\bi#1\ei{\begin{itemize}#1\end{itemize}}

\newcommand{\lefto}{\mathopen{}\left}

\DeclareMathOperator*{\argmax}{arg\;max}		
\DeclareMathOperator{\Exop}{\opE}			
\DeclareMathOperator{\Varop}{\opV\!\mathrm{ar}} 

\newcommand{\abs}[1]{\lefto\lvert#1\right\rvert}		

\safemath{\dirac}{\delta}					
\safemath{\krond}{\dirac}					

\safemath{\upto}{\uparrow}
\safemath{\downto}{\downarrow}
\safemath{\iu}{j}							
\safemath{\ev}{\lambda}						
\safemath{\hilseqspace}{l^{2}}				
\newcommand{\banachfunspace}[1]{\setL^{#1}}	
\safemath{\hilfunspace}{\banachfunspace{2}}	

\safemath{\SNR}{\textit{SNR}} 				
\safemath{\PAR}{\textit{PAR}} 				
\safemath{\No}{N_0}							
\safemath{\Es}{E_s}							
\safemath{\Eb}{E_b}							
\safemath{\EbNo}{\frac{\Eb}{\No}}
\safemath{\EsNo}{\frac{\Es}{\No}}

\DeclareMathOperator{\CHop}{\ensuremath{\opH}} 
\safemath{\tvir}{\rndh_{\CHop}}				
\safemath{\tvtf}{\rndl_{\CHop}}				
\safemath{\spf}{\rnds_{\CHop}}				
\safemath{\bff}{H_{\CHop}}					

\safemath{\ircf}{r_{h}}						
\safemath{\tftvcf}{r_{s}}					
\safemath{\tfcf}{r_{l}}						
\safemath{\bfcf}{r_{H}}						

\safemath{\tcorr}{c_h}						
\safemath{\scf}{c_{s}}						
\safemath{\tfcorr}{c_{l}}					
\safemath{\fcorr}{c_{H}}						

\safemath{\mi}{I}							
\safemath{\capacity}{C}						

\safemath{\normal}{\mathcal{N}}			
\safemath{\jpg}{\mathcal{CN}}			
\safemath{\mchain}{\leftrightarrow}		

\safemath{\dB}{\,\mathrm{dB}}
\safemath{\dBm}{\,\mathrm{dBm}}
\safemath{\Hz}{\,\mathrm{Hz}}
\safemath{\kHz}{\,\mathrm{kHz}}
\safemath{\MHz}{\,\mathrm{MHz}}
\safemath{\GHz}{\,\mathrm{GHz}}
\safemath{\s}{\,\mathrm{s}}
\safemath{\ms}{\,\mathrm{ms}}
\safemath{\mus}{\,\mathrm{\text{\textmu}s}}
\safemath{\ns}{\,\mathrm{ns}}
\safemath{\ps}{\,\mathrm{ps}}
\safemath{\meter}{\,\mathrm{m}}
\safemath{\mm}{\,\mathrm{mm}}
\safemath{\cm}{\,\mathrm{cm}}
\safemath{\m}{\,\mathrm{m}}
\safemath{\W}{\,\mathrm{W}}
\safemath{\mW}{\, \mathrm{mW}}
\safemath{\J}{\,\mathrm{J}}
\safemath{\K}{\,\mathrm{K}}
\safemath{\bit}{\,\mathrm{bit}}
\safemath{\nat}{\,\mathrm{nat}}

\safemath{\define}{\triangleq}			

\safemath{\equivalent}{\sim}
\safemath{\distas}{\sim}					
\safemath{\sdiff}{\Delta}				

\safemath{\reals}{\mathbb{R}}
\safemath{\positivereals}{\reals_{+}}
\safemath{\integers}{\mathbb{Z}}
\safemath{\posint}{\integers_{+}}
\safemath{\naturals}{\mathbb{N}}
\safemath{\posnaturals}{\naturals_{+}}
\safemath{\complexset}{\mathbb{C}}
\safemath{\rationals}{\mathbb{Q}}

\newcommand*{\fancyrefapplabelprefix}{app}		
\newcommand*{\fancyrefthmlabelprefix}{thm}		
\newcommand*{\fancyreflemlabelprefix}{lem}		
\newcommand*{\fancyrefcorlabelprefix}{cor}		
\newcommand*{\fancyrefdeflabelprefix}{def}		
\newcommand*{\fancyrefproplabelprefix}{prop}		
\newcommand*{\fancyrefexmpllabelprefix}{exmpl}
\newcommand*{\fancyrefalglabelprefix}{alg}		
\newcommand*{\fancyreftbllabelprefix}{tbl}		

\frefformat{vario}{\fancyrefseclabelprefix}{Section~#1}
\frefformat{vario}{\fancyrefthmlabelprefix}{Theorem~#1}
\frefformat{vario}{\fancyreftbllabelprefix}{Table~#1}
\frefformat{vario}{\fancyreflemlabelprefix}{Lemma~#1}
\frefformat{vario}{\fancyrefcorlabelprefix}{Corollary~#1}
\frefformat{vario}{\fancyrefdeflabelprefix}{Definition~#1}
\frefformat{vario}{\fancyreffiglabelprefix}{Fig.~#1}
\frefformat{vario}{\fancyrefapplabelprefix}{Appendix~#1}
\frefformat{vario}{\fancyrefeqlabelprefix}{(#1)}
\frefformat{vario}{\fancyrefproplabelprefix}{Proposition~#1}
\frefformat{vario}{\fancyrefexmpllabelprefix}{Example~#1}
\frefformat{vario}{\fancyrefalglabelprefix}{Algorithm~#1}

 \newtheorem{thm}{Theorem}
 
 \newtheorem{defi}{Definition}
 \newtheorem{lem}[thm]{Lemma}
 
\safemath{\dictab}{[\,\dicta\,\,\dictb\,]}

\safemath{\ysig}{\bmy}
\safemath{\ysighat}{\hat{\ysig}}
\safemath{\ysigdim}{M}
\safemath{\xsig}{\bmx}
\safemath{\xsigdim}{N}
\safemath{\nx}{n_x}
\safemath{\zsig}{\bmz}
\safemath{\zsigdim}{\ysigdim}
\safemath{\rsig}{\bmr}
\safemath{\Adict}{\bA}
\safemath{\Adicttilde}{\widetilde{\Adict}}
\safemath{\Adictdim}{\outputdim\times\xsigdim}
\safemath{\avec}{\bma}
\safemath{\avectilde}{\tilde{\avec}}
\safemath{\Bdict}{\bB}
\safemath{\Bdicttilde}{\widetilde{\Bdict}}
\safemath{\Cdict}{\bC}
\safemath{\cvec}{\bmc}
\safemath{\Ddict}{\bD}
\safemath{\Ddictdim}{\ysigdim\times\xsigdim}
\safemath{\dvec}{\bmd}
\safemath{\Ddicttilde}{\widetilde{\bD}}
\safemath{\Bonb}{\bB}
\safemath{\bvec}{\bmb}
\safemath{\Bonbdim}{\ysigdim\times\ysigdim}
\safemath{\noise}{\bmn}
\safemath{\noisedim}{\ysigim}
\safemath{\err}{\bme}
\safemath{\errdim}{\ysigdim}
\safemath{\errset}{\setE}
\safemath{\nerr}{n_e}
\safemath{\delop}{\bP_\errset}
\safemath{\delopc}{\bP_{{\errset}^c}}

\safemath{\cplxi}{\imath}
\safemath{\cplxj}{\jmath}

\safemath{\dict}{\matD}
\safemath{\inputdim}{N}		
\safemath{\outputdim}{M}		
\safemath{\sparsity}{S}	
\safemath{\inputdimA}{{N_a}}	
\safemath{\inputdimB}{{N_b}}	
\safemath{\elemA}{{n_a}}	
\safemath{\elemB}{{n_b}}	
\safemath{\resA}{\matR_a}	
\safemath{\resB}{\matR_b}	
\safemath{\subD}{\matS} 
\safemath{\subA}{\matS_a} 
\safemath{\subB}{\matS_b} 
\safemath{\dicta}{\matA} 	
\safemath{\dictb}{\matB} 	
\safemath{\hollowS}{H}
\safemath{\hollowA}{H_a}
\safemath{\hollowB}{H_b}
\safemath{\cross}{Z}
\safemath{\coh}{\mu_d}			
\safemath{\coha}{\mu_a}			
\safemath{\cohb}{\mu_b}			
\safemath{\mubs}{\nu}	
\safemath{\cohm}{\mu_m} 
\safemath{\dictset}{\setD}	
\safemath{\dictsetp}{\dictset(\coh,\coha,\cohb)}	
\safemath{\dictsetgen}{\dictset_\text{gen}}
\safemath{\dictsetgenp}{\dictsetgen(\coh)}
\safemath{\dictsetonb}{\dictset_\text{onb}}
\safemath{\dictsetonbp}{\dictsetonb(\coh)}

\safemath{\leftside}{U}
\safemath{\rightsideA}{R_a}
\safemath{\rightsideB}{R_b}

\safemath{\indexS}{\setI_S} 

\safemath{\na}{n_a}			
\safemath{\nb}{n_b}			
\safemath{\coeffa}{p_i}	
\safemath{\coeffb}{q_j}	
\safemath{\seta}{\setP}		
\safemath{\setb}{\setQ}     
\safemath{\setw}{\setW}	
\safemath{\setz}{\setZ}	
\safemath{\cola}{\veca}		
\safemath{\colb}{\vecb}		
\safemath{\cold}{\vecd}		
\safemath{\inputvec}{\vecx} 	
\safemath{\error}{\vece}	
\safemath{\noiseout}{\vecz} 	
\safemath{\inputvecel}{x}
\safemath{\inputveca}{\vecx_a}
\safemath{\inputvecb}{\vecx_b}
\safemath{\outputvec}{\vecy}	
\safemath{\lambdamin}{\lambda_{\mathrm{min}}}

\safemath{\elltwo}{\ell_2}
\safemath{\ellone}{\ell_1}
\safemath{\ellzero}{\ell_0}
\safemath{\ellinf}{\ell_\infty}
\safemath{\ellinftilde}{\ell_{\widetilde\infty}}
\safemath{\licard}{Z(\coh,\coha,\cohb)}
\safemath{\xsol}{\hat{x}}
\safemath{\xbord}{x_b}		
\safemath{\xstat}{x_s}		
\safemath{\xstatLone}{\tilde{x}_s}
\safemath{\order}{\mathcal{O}} 
\safemath{\scales}{\Theta} 
\safemath{\ones}{\mathbf{1}} 
\safemath{\zeroes}{\mathbf{0}} 
\safemath{\thlone}{\kappa(\coh,\cohb)} 
\safemath{\constoneA}{\delta} 
\safemath{\constoneB}{\epsilon} 
\safemath{\nlarge}{L}				   
\safemath{\sumlarge}{S_\nlarge}
\safemath{\maxlarger}{P_\nlarge}	   
\safemath{\Pzero}{\textrm{P0}}	
\safemath{\Pone}{\textrm{P1}}
\safemath{\vecfir}{\vecw}			 
\safemath{\vecsec}{\vecz}
\safemath{\elvecfir}{w}              
\safemath{\elvecsec}{z}				 
\safemath{\nlargefir}{n}
\safemath{\normout}{\gamma}
\safemath{\auxfun}{h}
\safemath{\supp}{\textrm{supp}}

\safemath{\indexa}{\ell}
\safemath{\indexb}{r}
\safemath{\indexc}{i}
\safemath{\indexd}{j}

\safemath{\project}{P}

\usepackage{color}

\newcommand{\shat}[1]{\hat\bms#1}
\newcommand{\shate}[1]{\hat{s}#1}

\newcommand{\srv}[1]{S#1}

\newcommand{\resid}[1]{\bmr#1}

\safemath{\LAMA}{\textrm{LAMA}}
\safemath{\MRT}{\textrm{MRT}}
\safemath{\betamax}{\beta^\textnormal{max}_\setO}
\safemath{\betamaxno}{\beta^\textnormal{max}}
\safemath{\betamin}{\beta^\textnormal{min}_\setO}
\safemath{\betaminno}{\beta^\textnormal{min}}

\safemath{\Nomin}{\No^\textnormal{min}(\beta)}
\safemath{\Nominnobeta}{\No^\textnormal{min}}
\safemath{\Nomax}{\No^\textnormal{max}(\beta)}
\safemath{\Nomaxnobeta}{\No^\textnormal{max}}

\IEEEoverridecommandlockouts 

\safemath{\MAP}{\textrm{MAP}}
\safemath{\IO}{\textrm{IO}}
\safemath{\JO}{\textrm{JO}}
\safemath{\Nopost}{N_{0}^\text{post}}
\safemath{\MT}{{M_\text{T}}}
\safemath{\MR}{{M_\text{R}}}
\safemath{\Tran}{\text{T}}
\safemath{\Herm}{\textnormal{H}}
\safemath{\row}{\text{r}}
\safemath{\col}{\text{c}}

\begin{document}

\title{Optimality of Large MIMO Detection via Approximate Message Passing}
\author{Charles Jeon, Ramina Ghods, Arian Maleki, and Christoph Studer
\thanks{An extended version of this paper including all proofs is in preparation \cite{JGMS2015}. 
}
\thanks{C. Jeon, R.~Ghods, and C.~Studer are with the School of Electrical and Computer Engineering, Cornell University, Ithaca, NY; e-mail: {jeon@csl.cornell.edu}, {rghods@csl.cornell.edu},  {studer@cornell.edu}.}
\thanks{A. Maleki is with Department of Statistics at Columbia University, New York City, NY; e-mail: {arian@stat.columbia.edu}.}
\thanks{
This work was supported in part by Xilinx Inc.~and by the US National Science
Foundation~(NSF) under grants ECCS-1408006 and CCF-1420328.
}}
\maketitle
\begin{abstract}
%
Optimal data detection in multiple-input multiple-output (MIMO) communication systems with a large number of antennas at both ends of the wireless link entails prohibitive computational complexity. 
In order to reduce the computational complexity, a variety of sub-optimal detection algorithms have been proposed in the literature. 
In this paper, we analyze the optimality of a novel data-detection method for large MIMO systems that relies on approximate message passing~(AMP).
We show that our algorithm, referred to as individually-optimal (IO) large-MIMO AMP (short IO-LAMA), is able to perform IO data detection given certain conditions on the~MIMO~system~and~the constellation~set~(e.g.,~QAM~or~PSK)~are~met. 
%
%
\end{abstract}




\section{Introduction}
\label{sec:intro}

We consider the problem of recovering the $\MT$-dimensional data vector $\vecs_0\in\setO^\MT$ from the noisy multiple-input multiple-output (MIMO) input-output relation \mbox{$\vecy=\bH\vecs_0+\bmn$}, by performing individually-optimal (IO) data detection \cite{guo2003multiuser,GV2005}
\begin{align*}
(\IO) \quad s_\ell^\IO & = \argmax_{\tilde s_\ell\in\setO} \,p\!\left(\tilde s_\ell \,|\, \bmy, \bH\right)\!.
\end{align*}
Here, $s_\ell^\IO$ denotes the $\ell$-th IO estimate, $\setO$ is a finite constellation (e.g., QAM or PSK), 
$p\!\left(\tilde s_\ell \,|\, \bmy, \bH\right)$ 
is a probability density function assuming i.i.d.\ zero-mean complex Gaussian noise for the vector \mbox{$\vecn\in\complexset^\MR$} with variance $\No$ per complex dimension,
$\MT$ and $\MR$ denotes the number of transmit and receive antennas, respectively, $\vecy\in\complexset^\MR$ is the receive vector, and \mbox{$\bH\in\complexset^{\MR\times\MT}$} is the (known)  MIMO system matrix.  
In what follows, we assume that the entries of the MIMO system matrix~$\bH$ are i.i.d.\ zero-mean complex Gaussian with variance $1/\MR$, and we define the so-called \emph{system ratio} as $\beta=\MT/\MR$. 

Although IO detection achieves the minimum symbol error-rate~\cite{V1998}, the combinatorial nature of the  (IO) problem \cite{guo2003multiuser,GV2005} requires prohibitive computational complexity, especially in large (or massive) MIMO systems \cite{V1998,SJSB11}. 
In order to enable data detection in such high-dimensional systems, a large number of low-complexity but  sub-optimal  algorithms have been proposed in the literature (see, e.g.,  \cite{indiachemp,CL2014,WYWDCS2014}).

\subsection{Contributions}

\begin{figure}[tp]
\centering
\subfigure[MIMO system with IO-\LAMA as the data detector.]{\includegraphics[width=0.84\columnwidth]{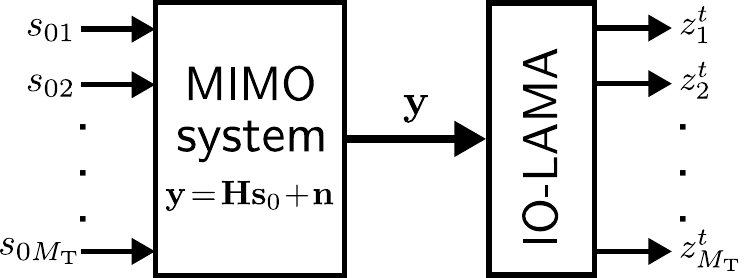}\label{fig:introfigurec}}\\
\subfigure[Equivalent decoupled system with effective noise variance $\sigma^2_t$.]{\includegraphics[width=0.84\columnwidth]{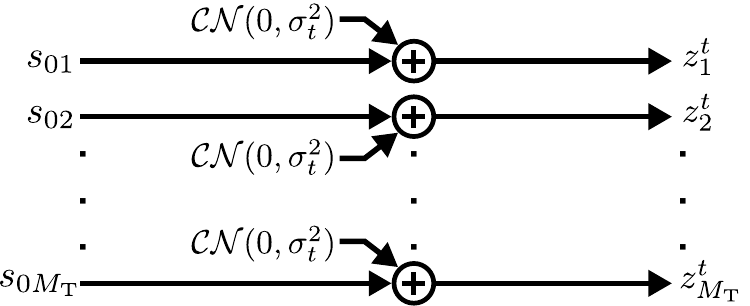}\label{fig:introfigured}}
%
\caption{IO-\LAMA decouples large MIMO systems (a) into a set of parallel and independent AWGN channels with equal noise variance;  (b) equivalent system in the large-system limit, i.e., for $\beta=\MT/\MR$ with $\MT\to\infty$.} 
\label{fig:introfigure2}
\vspace*{-0.3cm}
\end{figure}

In this paper, we propose and analyze a novel, computationally efficient data-detection algorithm, referred to as IO-\LAMA (short for IO \underline{la}rge \underline{M}IMO \underline{a}pproximate message passing). 
We show that IO-LAMA decouples the noisy MIMO system into a set of independent additive white Gaussian noise (AWGN) channels with equal signal-to-noise ratio (\SNR);  see \fref{fig:introfigure2} for an illustration of this decoupling property. 
The state-evolution~(SE) recursion of AMP enables us to track the effective noise variance  $\sigma_t^2$ of each decoupled AWGN channel at every algorithm iteration $t$.
Using these results, we provide precise conditions on the MIMO system matrix, the system ratio~$\beta$, the noise variance $\No$, and the modulation scheme for which IO-LAMA \emph{exactly} solves the (IO) problem. 




%

\subsection{Relevant Prior Art}

Initial results for IO data detection in large MIMO systems reach back to \cite{VS1999} where Verd\'u and Shamai analyzed the achievable rates under optimal data detection in randomly-spread CDMA systems. 
%
Tanaka \cite{TanakaCDMA} derived expressions for the error-rate performance and the multi-user efficiency for IO detection using the {replica method}.
While Tanaka's results were limited to BPSK constellations, Guo and Verd\'u extended his results to arbitrary discrete input distributions \cite{GV2005,GW2007}. 
All these results study the fundamental performance of IO data detection in the large-system limit, i.e., for $\beta=\MT/\MR$ with $\MT\to\infty$. Corresponding practical detection algorithms have been proposed for BPSK constellations \cite{K2003,CMT2004}---to the best of our knowledge, no computationally efficient algorithms for general constellation sets and complex-valued systems have been proposed in the open~literature. 

%

Our data-detection method, IO-LAMA, builds upon approximate message passing (AMP) \cite{donoho2009message,bayatimontanari,andreaGMCS}, which was initially developed for the recovery of sparse signals. 
AMP has been generalized to arbitrary signal priors in  \cite{DMM10a,DMM10b,malekiCAMP} and enables a precise performance analysis via the SE recursion \cite{donoho2009message,bayatimontanari}. 
%
%
Recently, AMP-related algorithms have been proposed for data detection 
\cite{NSE14,S2011,WKNLHG14}; these algorithms, however,  lack of a theoretical performance analysis.

\subsection{Notation}
Lowercase and uppercase boldface letters designate vectors and matrices, respectively. For a matrix $\bH$, we define its conjugate transpose to be $\bH^\Herm$. 
The $\ell$-th column 
of~$\bH$ is denoted by $\bmh_\ell^\col$.  
%
We use $\left\langle\cdot\right\rangle$ to write $\left\langle \bmx \right\rangle = \frac{1}{N}\sum_{k=1}^N x_k$. A multivariate complex-valued Gaussian probability density function (pdf) is denoted by $\setC\setN(\bmm,\bK)$, where~$\bmm$ is the mean vector and $\bK$ the covariance matrix. $\Exop_X\!\left[\cdot\right]$ and $\Varop_X\!\left[\cdot\right]$ denotes the expectation and variance operator with respect to the pdf of the random variable~$X$, respectively.

\section{IO-LAMA: Large-MIMO Detection using AMP}\label{sec:DAMP}

We now present IO-\LAMA and the SE recursion, which is used in \fref{sec:SE_MI_ER} for our optimality analysis. 

\subsection{The IO-LAMA Algorithm}

We assume that the transmit symbols $s_{\ell}$, $\ell=1,\ldots,\MT$, of the transmit data vector $\bms$ are taken from a finite set $\setO=\{a_j : j=1,\ldots,|\setO|\}$ with constellation points~$a_j$ chosen, e.g., from a QAM or PSK alphabet.
We assume an i.i.d.\ prior $p(\vecs)=\prod_{\ell=1}^{\MT}p(s_{\ell})$, with the following distribution for each transmit symbol $s_{\ell}$: %
\begin{align} \label{eq:prior}
\textstyle p(s_{\ell}) = \sum_{a\in \setO}p_a\delta(s_{\ell} - a).
\end{align}
Here, $p_{a}$ designates the (known) prior probability of each constellation point $a\in\setO$ and $\delta(\cdot)$ is the Dirac delta function; for uniform priors, we have $p_{a}=|\setO|^{-1}$.

The IO-\LAMA algorithm summarized below is obtained by using the prior distribution in \fref{eq:prior} within complex Bayesian AMP. A detailed derivation of the algorithm is given in \cite{JGMS2015}.
\newtheorem{alg}{Algorithm}
\begin{alg} Initialize $\hat{s}^1_\ell=\Exop_S[S]$ for $\ell=1,\ldots,\MT$, \mbox{$\resid^1= \bmy$}, and $\tau^1=\beta\Varop_S[S]/\No$. Then, for every IO-LAMA iteration $t=1,2,\ldots,$ compute the following steps: 
\begin{align*}
\bmz^{t}&=\hat\bms^t+\bH^\Herm\bmr^t\\
\shat^{t+1} &= \textstyle \mathsf{F}(\bmz^{t},\No(1+\tau^t))\\
\tau^{t+1} &= \textstyle \frac{\beta}{\No}\!\left\langle\mathsf{G}(\bmz^{t},\No(1+\tau^t))\right\rangle\\
\resid^{t+1}  &= \textstyle  \bmy-\bH\shat^{t+1}+\frac{\tau^{t+1}}{1+\tau^t}\resid^{t}.
\end{align*}
The functions $\mathsf{F}(s_{\ell},\tau)$ and $\mathsf{G}(s_{\ell},\tau)$ correspond to the message mean and variance, and are computed as follows:
\begin{align}\label{eq:Ffunc}
\mathsf{F}(\hat{s}_{\ell},\tau) &= \textstyle
\int_{s_\ell} s_{\ell}f(s_{\ell}\vert \shate_{\ell},\tau)\mathrm{d}s_{\ell}\\\notag
\mathsf{G}(\shate_\ell,\tau)&= \textstyle \int_{s_\ell}\abs{s_{\ell}}^2f(s_{\ell}\vert \shate_{\ell},\tau)\mathrm{d}s_{\ell}- \abs{\mathsf{F}(\shate_{\ell},\tau)}^2.
\end{align}
Here, $f(s_{\ell}\vert\hat{s}_{\ell},\tau)$ is the posterior pdf defined by 
$f(s_{\ell}\vert\hat{s}_{\ell},\tau)=\frac{1}{Z}p(\hat{s}_{\ell}\vert s_{\ell},\tau)p(s_\ell)$
with $p(\hat{s}_{\ell}\vert s_{\ell},\tau)\sim\setC\setN(s_{\ell},\tau)$ and a normalization constant $Z$. Both functions $
\mathsf{F}(\hat{s}_{\ell},\tau)$ and $
\mathsf{G}(\shate_\ell,\tau)$ operate element-wise on vectors. 
\end{alg}

In order to analyze the performance of  IO-\LAMA in the large-system limit, we next summarize the SE recursion.
The SE recursion in the following theorem enables us to track the effective noise variance $\sigma_{t}^2$ for the decoupled MIMO system for every iteration~$t$  (cf.~\fref{fig:introfigure2}), which is key for the optimality analysis in \fref{sec:SE_MI_ER}. 
A detailed derivation is given in \cite{JGMS2015}.

\begin{thm}\label{thm:CSE} 
Fix the system ratio $\beta=\MT/\MR$ and the constellation set $\setO$, and let $\MT\rightarrow\infty$. Initialize $\sigma_1^2=\No+\beta\Varop_S[S]$. Then, the effective noise variance $\sigma_{t}^2$ of IO-LAMA at iteration~$t$  is given by the following recursion:
\begin{align}
\sigma_{t}^2 &  
 = \No +\beta\Psi(\sigma_{t-1}^2). \label{eq:SErecursion}
\end{align}The so-called mean-squared error (MSE) function is defined~by
\begin{align*}
\textstyle \Psi(\sigma_{t-1}^2) = \Exop_{\srv,Z}\!\left[\abs{ \mathsf{F}\!\left(\srv + \sigma_{t-1} Z,\sigma_{t-1}^2\right)\! - \srv}^2 \right]\!,
\end{align*}
where $\mathsf{F}$ is given in \fref{eq:Ffunc} and $Z\sim\setC\setN(0,1)$. 
\end{thm}

\subsection{IO-\LAMA Decouples Large MIMO Systems}\label{sec:LAMAdecouple}

%
%
%
%

%
In the large-system limit and for every iteration $t$, IO-LAMA computes the marginal distribution of $s_\ell$, $\ell=1,\ldots,\MT$, which corresponds to a Gaussian distribution centered around the original signal $s_{0\ell}$ with variance $\sigma_{t}^2$. 
These properties follow from \cite[Sec. 6]{andreaGMCS}, which shows that $\bmz^t=\shat^t+\bH^\Herm\bmr^t$ is distributed according to $\setC\setN(\bms_0,\sigma_t^2\bI_\MT)$.
Hence, the input-output relation for each transmit stream $z_\ell^t = \shate_\ell^t+ (\bmh_\ell^\col)^\Herm\bmr_\ell^t$ is equivalent to the following single-stream AWGN channel:
\begin{align*}
z_\ell^t = s_{0\ell} + n_\ell^t.
\end{align*}
Here, $s_{0\ell}$ is the $\ell$-th original transmitted signal and $n_\ell^t$ is AWGN with 
variance $\sigma^2_t$ per complex entry. 
%
%
As a consequence, IO-LAMA decouples the MIMO system into $\MT$ parallel and independent AWGN channels with equal noise variance $\sigma^2_t$ in the large-MIMO limit; see~\fref{fig:introfigured} for an illustration. 

\section{Optimality of IO-LAMA}\label{sec:SE_MI_ER}

We now provide conditions for which IO-\LAMA \emph{exactly} solves the (IO) problem. 
%

\subsection{Fixed points of IO-\LAMA's State Evolution}
\label{sec:optimality}



For $t\to\infty$, the SE recursion in \fref{thm:CSE} converges to the following fixed-point equation~\cite{bayatimontanari,JGMS2015}:
\begin{align}\label{eq:fixed_pt}
\sigma_\IO^2 = \No + \beta \Psi(\sigma_\IO^2),
\end{align}
which coincides with the ``fixed-point equation'' developed for IO detection by Guo and Verd\'u using the replica method in~\cite[Eq. (34)]{GV2005}. 
%
%
We note that \eqref{eq:fixed_pt} may have multiple fixed-point solutions. In the case of such non-unique fixed points, Guo and Verd\'u choose the solution that minimizes the ``free energy'' \cite[Sec. 2-D]{GV2005}, whereas IO-\LAMA converges, in general, to the fixed-point solution with the largest effective noise variance~$\sigma^2$.
We note that if the fixed-point solution to \fref{eq:fixed_pt} is unique, then IO-\LAMA recovers the solution with minimal effective noise variance $\sigma^2$ and thus, performs IO detection. 
However, if there are multiple fixed-points solutions to \fref{eq:fixed_pt}, IO-\LAMA is, in general, sub-optimal and does not necessarily converge to the fixed-point solution with the minimal ``free energy.''\footnote{Convergence to another fixed-point solution is possible if IO-LAMA is initialized sufficiently close to such a fixed point; see \cite{JGMS2015,ZMWL2015} for the details.}
We next provide conditions for which there is exactly one (unique) fixed point with minimum effective noise variance~$\sigma^2$ and---as a consequence---IO-\LAMA solves the (IO) problem.

\subsection{Exact Recovery Thresholds (ERTs)}\label{sec:ERTs}

We start by analyzing IO-\LAMA in the noiseless setting.
We provide conditions on the system ratio $\beta$ and the constellation set $\setO$, which guarantee exact recovery of an unknown transmit signal $\vecs_0\in\setO^\MT$ in the large-system limit, i.e., $\beta$ is fixed and $\MT\to\infty$. In particular, we show that if $\beta<\betamax$, where $\betamax$ is the so-called \emph{exact recovery threshold~(ERT)}, then IO-\LAMA perfectly recovers $\vecs_0$; for $\beta\geq\betamax$, perfect recovery is not guaranteed, in general.\footnote{We assume the initialization in Algorithm 1. IO-LAMA may recover the original signal for $\beta\geq\betamax$ if initialized appropriately; see, e.g., \cite{ZMWL2015}.} 
To make this behavior explicit, we need the following technical result; the proof is given~in~\fref{app:DAMPsolvability}.
%
\begin{lem}\label{lem:DAMPsolvability} Fix the constellation set $\setO$. If $\Varop_S[S]$ is finite, then there exists a non-negative gap $\sigma^2 - \Psi(\sigma^2)\geq 0 $ with equality if and only if $\sigma^2 = 0$. As $\sigma^2\rightarrow 0$, the MSE $\Psi(\sigma^2)\rightarrow 0$ and as $\sigma^2\rightarrow\infty$, MSE $\Psi(\sigma^2)\rightarrow \Varop_S[S]$.
\end{lem}

%

%
For all \mbox{$\sigma^2>0$}, \fref{lem:DAMPsolvability} guarantees that \mbox{$\Psi(\sigma^2)<\sigma^2$}. 
Suppose that for some $\beta>1$, \mbox{$\beta\Psi(\sigma^2)<\sigma^2$} also holds for all $\sigma^2>0$.
Then, as long as $\beta>1$ is not too large to also ensure \mbox{$\beta\Psi(\sigma^2)<\sigma^2$} for all \mbox{$\sigma^2>0$}, there will only be a \emph{single} fixed point at \mbox{$\sigma^2=0$}. Therefore, \LAMA can still perfectly recover the original signal $\vecs_0$ by \fref{thm:CSE} since \mbox{$\Psi(\sigma^2)=0$}. 
Leveraging the gap between $\Psi(\sigma^2)$ and $\sigma^2$ will allow us to find the exact recovery threshold (ERT) of \LAMA for values of $\beta>1$.
For the fixed (discrete) constellation set $\setO$, the largest $\beta$ that ensures $\beta\Psi(\sigma^2)<\sigma^2$ is precisely the ERT defined next.

\begin{defi} \label{def:maxbeta}Fix $\setO$ and let \mbox{$\No=0$}. Then, the exact recovery threshold (ERT) that enables perfect recovery of the original signal $\bms_0$ using IO-\LAMA is given by
\begin{align}\label{eq:beta_recover}
\betamax &= \min_{\sigma^2>0}\!\left\{\!\left(\frac{\Psi(\sigma^2)}{\sigma^2}\right)^{\!\!-1}\right\}\!.
\end{align}
\end{defi}
With \fref{def:maxbeta}, we state \fref{thm:recovery}, which establishes optimality in the noiseless case; the proof is given in \fref{app:recovery}.
\begin{thm} \label{thm:recovery} Let $\No=0$ and fix a discrete set $\setO$. 
If \mbox{$\beta<\betamax$}, then IO-\LAMA perfectly recovers the original signal $\bms_0$ from $ \mbox{$\vecy=\bH\vecs_0+\bmn$}$ in the large system limit.
\end{thm}
Note that for a given constellation set $\setO$, the ERT $\betamax$ can be computed numerically using \fref{eq:beta_recover}.
%
Furthermore, the signal variance, $\Varop_S[S]$, has no impact on the ERT as the MSE function $\Psi(\sigma^2)$ and $\sigma^2$ scale linearly with $\Varop_S[S]$.
Table~\ref{tbl:exact_recovery} summarizes ERTs $\betamax$ for common QAM and PSK constellation sets. 

\begin{table}
\centering
\caption{ERTs $\beta_\setO^\text{max}$, MRTs $\beta_\setO^\text{min}$, and critical noise levels $\Nominnobeta(\betamin)$ and $\Nomaxnobeta(\betamax)$
of IO-\LAMA for common constellation sets}
\begin{tabular}{lccccc}
\toprule
Constellation & $\betamin$ & $\Nominnobeta(\betamin)$
& $\betamax$
& $\Nomaxnobeta(\betamax)$
\\
\midrule
BPSK  & 2.9505 & $2.999\cdot 10^{-1}$ & 4.1709& $2.432\cdot 10^{-1}$ \\
QPSK & 1.4752 & $1.499\cdot 10^{-1}$ & 2.0855& $1.216\cdot 10^{-1}$\\
%
16-QAM  & 0.9830 & $3.000\cdot 10^{-2}$ & 1.3629 & $2.454\cdot 10^{-2}$\\
64-QAM & 0.8424 & $7.144\cdot 10^{-3}$& 1.1573 & $5.868\cdot 10^{-3}$\\
8-PSK& 1.4576 & $4.440\cdot 10^{-2}$ & 1.8038&$3.826\cdot 10^{-2}$
\\
16-PSK  & 1.4728 & $1.143\cdot 10^{-2}$ & 1.8005& $9.953\cdot 10^{-3}$\\
\bottomrule
\end{tabular}
\label{tbl:exact_recovery}
\vspace{-0.2cm}
\end{table}

\subsection{Optimality Conditions for IO-\LAMA}\label{sec:LAMA_optimal}

\begin{figure*}[tp]
\centering
\subfigure[$\beta\leq\betamin$: IO-\LAMA always converges to the unique, optimal fixed point (FP)~irrespective~of~$\No$. ]{\includegraphics[width=0.32\textwidth]{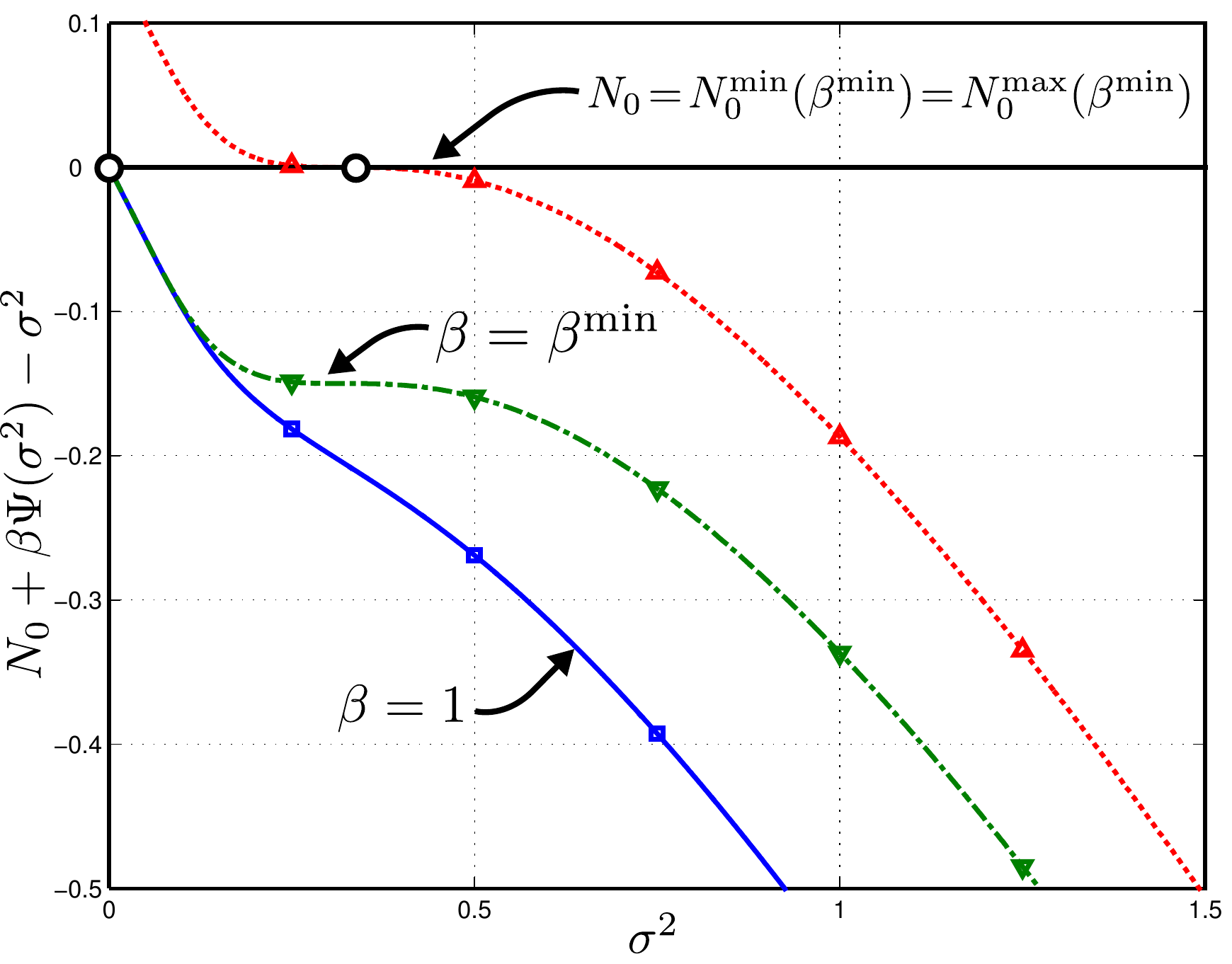}\label{fig:SE_QPSK1}}
\hspace{0.4mm}
\subfigure[$\beta\in(\betamin,\betamax)$: IO-\LAMA converges to the optimal FP if \mbox{$\No<\Nomin$} or \mbox{$\No>\Nomax$}.]{\includegraphics[width=0.32\textwidth]{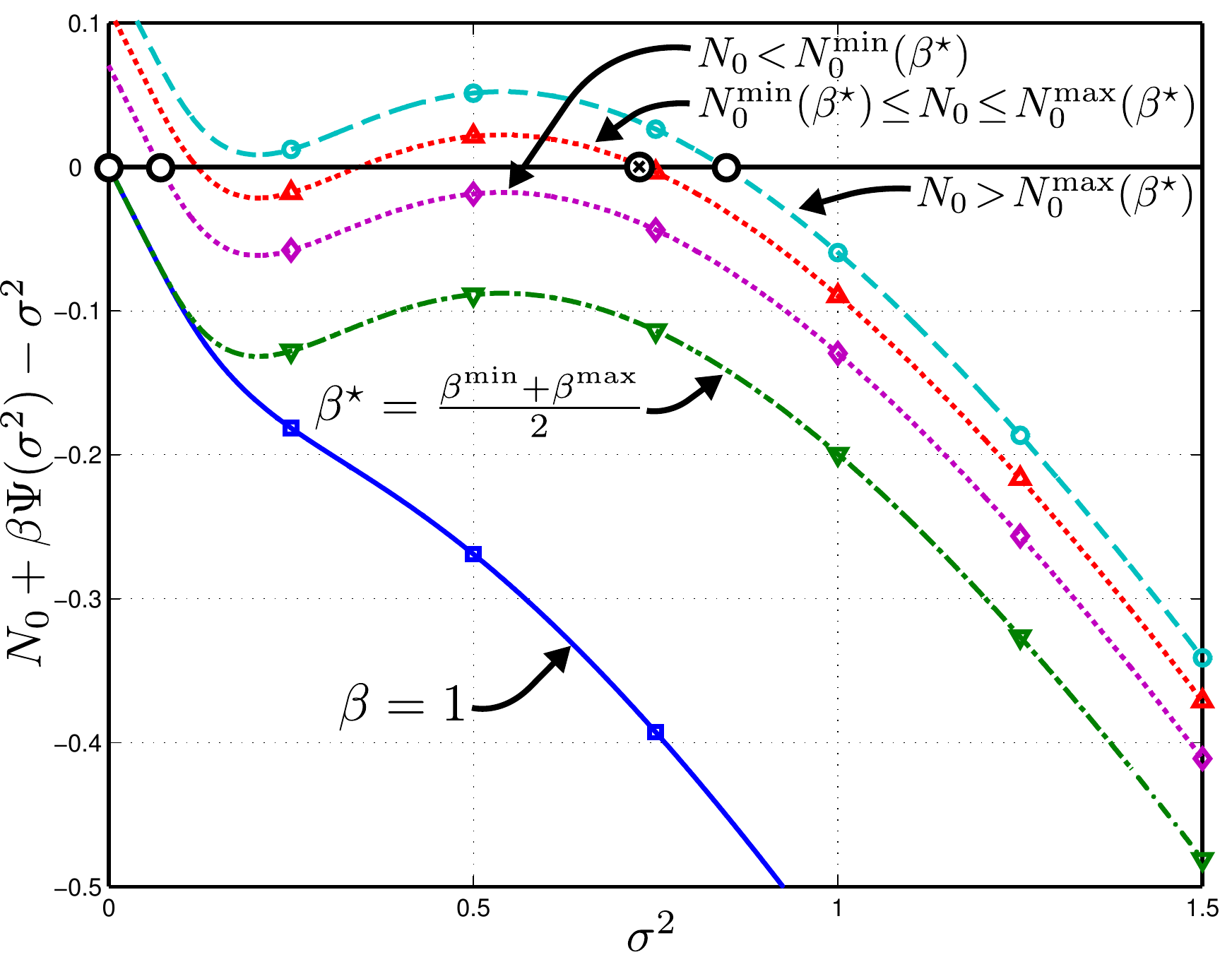}\label{fig:SE_QPSK2}}
\hspace{0.4mm}
\subfigure[$\beta\geq\betamax$: IO-\LAMA converges to the optimal fixed point if $\No>\Nomax$.]{\includegraphics[width=0.32\textwidth]{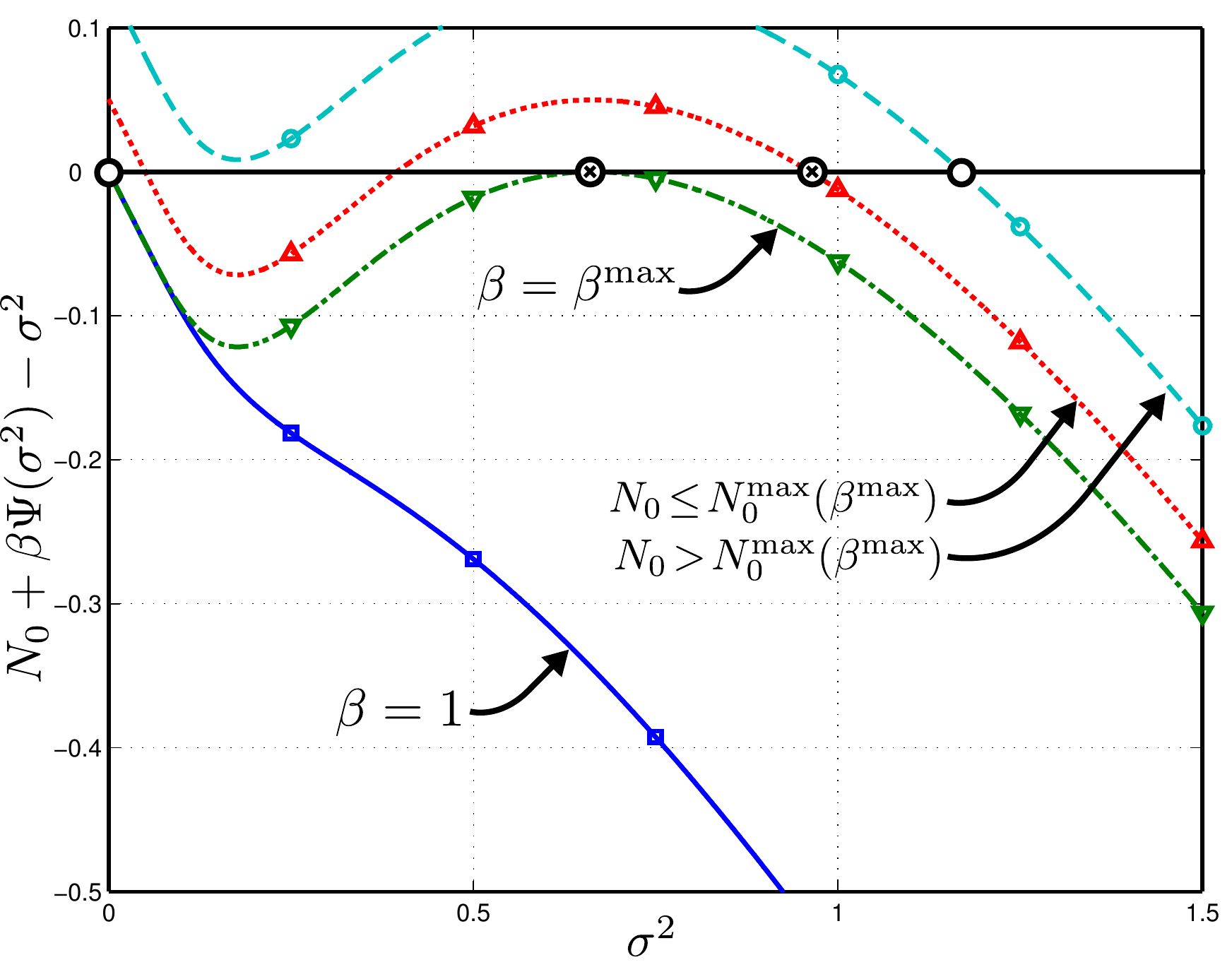}\label{fig:SE_QPSK3}}
\caption{
Plot of the function \fref{eq:plotfixedfunction} for three regimes (a) $\beta\leq\betamin$, (b) $\beta\in(\betamin,\betamax)$, and (c) $\beta\geq\betamax$ for  QPSK modulation, uniform priors, and $\Varop_S[S]=\Es=1$. 
%
The optimal fixed points are designated by $\circ$; suboptimal fixed points are designated by $\otimes$.
%
%
%
%
}
\label{fig:SE_QPSK}
\vspace{-0.2cm}
\end{figure*}

We now study the optimality of IO-\LAMA in the presence of noise, where  \emph{exact} recovery is no longer guaranteed. In particular, we provide conditions for which IO-\LAMA converges to the fixed point with minimal effective noise variance $\sigma^2$, which corresponds to solving the (IO) problem. 
Note that such a minimum free-energy solution is also the fixed point for the IO detector in \cite[Eq. (34)]{GV2005}.
We call the fixed point with minimum effective noise variance \emph{optimal fixed point}; other fixed points are called \emph{suboptimal fixed points}. 

\begin{table}
\centering
\caption{Summary of (sub-)optimality regimes of  IO-\LAMA}
\begin{tabular}{lccc}
\toprule
&$\beta\leq\betamin$\hspace*{-1mm}&$\betamin\!<\!\beta\!<\!\betamax$\hspace*{-1mm}&$\betamax\leq\beta$\\
\midrule
\!\!\!\hspace{14.3mm}$\No<\Nomin$\hspace*{-3mm}&\em optimal&\em optimal& suboptimal\\
\!\!\!$\Nomin\leq\No\leq\Nomax$\hspace*{-3mm}
&\em optimal& (sub-)optimal\tablefootnote{For certain constellation sets (e.g., 16-PSK), there exist sub-intervals in $[\Nomin,\!\Nomax]$ where IO-\LAMA is still optimal; see \cite{JGMS2015} for the details.} &suboptimal\\
\!\!\!$\Nomax<\No$&\em optimal&\em optimal&\em optimal\\
\bottomrule
\end{tabular} 
\label{tbl:IOLAMAoptimal_reg}
\vspace{-0.2cm}
\end{table}

We identify three different operation regimes for IO-\LAMA depending on the system ratio $\beta$  (see \fref{tbl:IOLAMAoptimal_reg}).
To make these three regimes explicit, we need the following definition.


\begin{defi}\label{def:twobeta}Fix the constellation set $\setO$. Then, the minimum recovery threshold (MRT) $\betamin$ is defined by
\begin{align}\label{eq:beta_badstate}
\betamin&=\min_{\sigma^2>0}\!\left\{\!\left(\frac{\textnormal{d}\Psi(\sigma^2)}{\textnormal{d}\sigma^2}\right)^{\!\!-1}\right\}\!.
\end{align}
\end{defi}
The definition of MRT shows that for all system ratios \mbox{$\beta\leq\betamin$}, the fixed point of \fref{eq:fixed_pt} is unique. 
The following lemma establishes a fundamental relationship between MRT and ERT; the proof is given in \fref{app:MRTandERT}.

\begin{lem} The MRT never exceeds the ERT. \label{lem:MRTandERT}
\end{lem} 

We next define the minimum critical and maximum guaranteed noise variance, \Nomin and \Nomax, that determine boundaries for the optimality regimes when $\beta>\betamin$. 

\begin{defi} \label{def:critical_SNR}Fix \mbox{$\beta\in(\betamin,\betamax)$}. Then, the \emph{minimum critical} noise $\Nomin$ that ensures convergence to the optimal fixed point is defined~by
\vspace*{-0.2cm}
\begin{align*}
\Nomin &=  \min_{\sigma^2>0} \!\left\{\sigma^2-\beta\Psi(\sigma^2):\beta\frac{\textnormal{d}\Psi(\sigma^2)}{\textnormal{d}\sigma^2}=1\right\}\!.
\end{align*}
\end{defi}

\begin{defi}\label{def:max_SNR}Fix \mbox{$\beta>\betamin$}. Then, the \emph{maximum guaranteed} noise $\Nomax$ that ensures convergence to the optimal fixed point is defined~by
\begin{align*}
\Nomax &= \max_{\sigma^2>0}\!\left\{\sigma^2 - \beta\Psi(\sigma^2):\beta\frac{\textnormal{d}\Psi(\sigma^2)}{\textnormal{d}\sigma^2} = 1
\right\}\!.
\end{align*}
\end{defi}
We recall that all the zero crossings of the function
\begin{align}\label{eq:plotfixedfunction}
g(\sigma^2,\beta,\No)_\setO=\No+\beta\Psi(\sigma^2)-\sigma^2
\end{align}
correspond to all fixed points of the SE recursion of IO-\LAMA; we use this function to study the algorithm's optimality.

Figure \ref{fig:SE_QPSK} illustrates our optimality analysis for a large-MIMO system with QPSK constellations. 
We show \fref{eq:plotfixedfunction} depending on the effective noise variance $\sigma^2$ and for different system ratios~$\beta$. The regimes $\beta\leq\betamin$, $\beta\in(\betamin,\betamax)$, and $\beta\geq\betamax$ are shown in \fref{fig:SE_QPSK1}, \fref{fig:SE_QPSK2}, and \fref{fig:SE_QPSK3}, respectively.
The special case for $\beta=1$ with $\No=0$ corresponds to the solid blue line, along with the corresponding (unique) fixed point at the origin. In the following three paragraphs, we discuss the three operation regimes of IO-LAMA in detail.


%
\subsubsection*{(i) \mbox{$\beta\leq\betamin$}}
In this region, the SE recursion of IO-\LAMA always converges to the unique, optimal fixed point. 
For \mbox{$\beta<\betamin$},
the slope of \fref{eq:plotfixedfunction} for all $\sigma^2$ is strictly-negative. Hence, as \fref{eq:plotfixedfunction} is always decreasing, there exists exactly one unique fixed point of the SE recursion regardless of the noise variance $\No$. 
Thus, IO-LAMA converges to the optimal fixed point and consequently, solves the (IO) problem. 

We emphasize that we still obtain exactly one fixed point even when $\beta$ is equal to the MRT.
Since $\beta=\betamin$, there exists at least one $\sigma_\star^2$ that satisfies $\betamin\frac{\textnormal{d}}{\textnormal{d}\sigma^2}\Psi(\sigma^2)\big\vert_{\sigma^2=\sigma_\star^2}=1$. 
By definition of $\betamin$, \fref{eq:plotfixedfunction} at $\sigma_\star^2$ implies that $\sigma_\star^2$ is a saddle-point, so \fref{eq:plotfixedfunction} has exactly one zero at $\sigma_\star^2$.
We observe that if $\sigma_\star^2$ is unique, then $\Nominnobeta(\betamin) = \Nomaxnobeta(\betamin)$.
For all other $\sigma^2\neq\sigma_\star^2$, the construction of $\sigma_\star^2$ implies that $\betamin\frac{\textnormal{d}}{\textnormal{d}\sigma^2}\Psi(\sigma^2)<1$, so the fixed point of \fref{eq:plotfixedfunction} remains to be unique. 

The green, dash-dotted and red, dotted line in \fref{fig:SE_QPSK1} shows \fref{eq:plotfixedfunction} for \mbox{$\beta=\betamin$} with \mbox{$\No=0$} and \mbox{$\No=\Nominnobeta(\betamin)=\Nomaxnobeta(\betamin)$}, respectively. In both cases, we see that the SE recursion of IO-\LAMA converges to the unique fixed point.

\subsubsection*{(ii) \mbox{$\betamin<\beta<\betamax$}}
In this region, the SE recursion of IO-\LAMA converges to the unique, optimal fixed point if \mbox{$\No<\Nomin$} or \mbox{$\No>\Nomax$}.
%

The green, dash-dotted line, cyan, dashed line, and magenta, dotted line in \fref{fig:SE_QPSK2} shows \fref{eq:plotfixedfunction} for $\beta^\star=(\betamin+\betamax)/{2}$ with \mbox{$\No=0$}, \mbox{$\No>\Nomaxnobeta(\beta^\star)$} and \mbox{$\No<\Nominnobeta(\beta^\star)$}, respectively. We note that for the three cases, the fixed point is unique, labeled in \fref{fig:SE_QPSK2} by a circle. 
On the other hand, the red, dotted line in \fref{fig:SE_QPSK2} shows \fref{eq:plotfixedfunction} with~$\beta^\star$ under noise $\No\in[\Nominnobeta(\beta^\star),\Nomaxnobeta(\beta^\star)]$.  
In this case, however, we observe that SE recursion of IO-\LAMA converges to the rightmost suboptimal fixed point labeled by the crossed circle $\otimes$. Hence, IO-\LAMA does not, in general, solve the (IO) problem when $\Nomin\leq\No\leq\Nomax$.

\subsubsection*{(iii) \mbox{$\beta\geq\betamax$}}

In this region, the SE recursion of IO-\LAMA converges to the unique, optimal fixed point when \mbox{$\No>\Nomax$}.
%
%
As $\beta\rightarrow\betamax$, the low noise \mbox{$\No<\Nomin$} (or high \SNR) region of optimality disappears because $\Nomin\rightarrow0$ as $\beta\rightarrow\betamax$ from \fref{eq:beta_recover}.

The green, dash-dotted line and red, dotted line in \fref{fig:SE_QPSK3} shows \fref{eq:plotfixedfunction} for $\beta=\betamax$ with $\No=0$ and $0<\No\leq\Nomax$, respectively. We observe that the SE recursion of IO-\LAMA converges to the suboptimal fixed point when $\beta=\betamax$ even with $\No=0$. On the other hand, the cyan, dashed line refers to~\fref{eq:plotfixedfunction} for $\beta=\betamax$ with $\No>\Nomax$. 
While the noiseless case resulted the SE recursion of IO-\LAMA to converge to the suboptimal fixed point, we observe that for strong noise (or equivalently low \SNR), the SE recursion of IO-\LAMA actually recovers the IO solution.
Therefore, when \mbox{$\beta\geq\betamax$}, IO-\LAMA solves the (IO) problem when the noise is greater than the maximum guaranteed~noise~$\Nomax$.

As a final remark, we note that the ERT $\betamax$ and MRT $\betamin$ in \fref{tbl:exact_recovery} do not depend on $\Varop_S[S]$; the critical noise levels $\Nomin$ and $\Nomax$, however, depend on $\Varop_S[S]$.

\subsection{ERT, MRT, and Critical Noise Levels}\label{sec:ERTMRT}

The ERT, MRT, as well as $\Nomin$ and $\Nomax$ for common constellations are summarized in \fref{tbl:exact_recovery}.
We assume equally likely priors with the transmit signal normalized to \mbox{$E_s=\Varop_S[S]=1$}.\footnote{The critical noise levels depend linearly on $E_s$. Hence, we assume that $E_s=1$ without loss of generality.} 
We note that the calculations of ERT and MRT for the simplest case of~BPSK constellations involve computations of a logistic-normal integral for which no closed-form expression is known \cite{D2005}.
Consequently, the following results were obtained via numerical integration for computing the MSE function $\Psi(\sigma^2)$. 
%
%
%
%
As noted in \fref{tbl:exact_recovery} for a QPSK system under complex-valued noise, the ERT is \mbox{$\betamaxno_\text{QPSK}\approx2.0855$}, and the MRT is given as \mbox{$\betaminno_\text{QPSK}\approx1.4752$}.
%

The MRTs for 16-QAM and 64-QAM indicate that small system ratios $\beta<1$ are required to always guarantee that IO-\LAMA solves the (IO) problem in the presence of noise. 
For instance, we require \mbox{$\beta\leq\beta^\text{min}_\text{64-QAM}\approx0.8424$}, i.e. $\MT\leq0.8424\MR$,  to ensure that IO-\LAMA solves the IO problem for 64-QAM in the large system limit. As \mbox{$\beta\rightarrow\beta^\text{max}_\text{64-QAM}\approx1.1573$}, IO-\LAMA is only optimal for \mbox{$\No>\Nomaxnobeta(\beta^\text{max}_\text{64-QAM})\approx5.868\cdot10^{-3}$}.
%
From \fref{tbl:exact_recovery}, we see that IO-LAMA is a suitable candidate algorithm for the detection of higher-order QAM constellations in massive multi-user MIMO systems as one typically assumes \mbox{$\MR\gg\MT$}\cite{LETM2014}.  

\section{Conclusions}
We have presented the IO-\LAMA algorithm along with the state-evolution recursion.
Using these results, we have established conditions on the MIMO system matrix, the noise variance $\No$, and the constellation set for which IO-\LAMA exactly solves the (IO) problem. 
While the presented results are exclusively for the large-system limit, our own simulations indicate that IO-LAMA achieves near-optimal performance in realistic, finite-dimensional systems; see \cite{JGMS2015} for more details.

\appendices

\section{Proof of \fref{lem:DAMPsolvability}}\label{app:DAMPsolvability}
%
Since the variance of $S$ is finite, denote $\Varop_S[S]=\sigma_s^2$. By \cite[Prop.~5]{GWSS2011}, we have the following upper bound:
\begin{align}\label{eq:MSIbeta1}
\Psi(\sigma^2)&\leq\frac{\sigma_s^2}{\sigma_s^2+\sigma^2}\sigma^2= \frac{1}{1+\sigma^2/\sigma_s^2}\sigma^2.
\end{align}
Here, equality holds for all $\sigma^2$ if and only if $S$ is complex normal with variance $\sigma_s^2$ \cite{GWSS2011}. Note that if $\sigma^2=0$, then \fref{eq:MSIbeta1} is achieved for any $\sigma_s^2$. If $\sigma^2>0$, then $\Psi(\sigma^2)<\sigma^2$ by \fref{eq:MSIbeta1}.

The first part follows directly from \fref{eq:MSIbeta1} as $\Psi(\sigma^2)$ is non-negative. The second part requires one to realize that $\sigma^2\rightarrow\infty$ also implies $\mathsf{F}(\cdot,\sigma^2)\rightarrow \sum_{a\in\setO}a p_a = \Exop_S[S]$, and hence,
\begin{align*}
\lim_{\sigma^2\rightarrow\infty}\!\Psi(\sigma^2) \rightarrow\Exop_{S}\!\left[
\abs{
	S - \Exop_S[S]
}^2
\right] = \Varop_S[S].
\end{align*}


\section{Proof of \fref{thm:recovery}}\label{app:recovery}

We assume the initialization in Algorithm 1. Since $\No=0$, if \LAMA perfectly recovers the original signal $\bms_0$, then the fixed point in \fref{eq:fixed_pt} is unique at \mbox{$\sigma^2=0$}.
This happens if the system ratio is strictly less than the ERT $\betamax$ because otherwise, i.e., $\beta\geq\betamax$, there exists a non-unique fixed point to \fref{eq:fixed_pt} for some $\sigma^2>0$ by \fref{def:maxbeta}.
\section{Proof of \fref{lem:MRTandERT}}\label{app:MRTandERT}
We show that under a fixed constellation set $\setO$, $\betamin\leq\betamax$. The proof is straightforward as,
\begin{align*}
\betamin&\textstyle\stackrel{(a)}{=}\underset{\sigma^2>0}{\min}\!\left\{\!\left(\frac{\textnormal{d}\Psi(\sigma^2)}{\textnormal{d}\sigma^2}\right)^{\!\!-1}\right\}\!\leq
\!\left(\frac{\textnormal{d}\Psi(\sigma^2)}{\textnormal{d}\sigma^2}\right)^{\!\!-1}\Big\vert_{\sigma^2=\beta_\setO^\textnormal{max}\Psi(\sigma^2)}\\
&\textstyle\stackrel{(b)}{=}\!\left(\frac{1}{\betamax}\right)^{\!\!-1}=\betamax,
\end{align*}
where (a) and (b) follow from the MRT and ERT definitions.

\bibliographystyle{IEEEtran}
\bibliography{confs-jrnls,publishers,VIP}

\end{document}